\documentclass[aps,prl,twocolumn,showkeys,showpacs,superscriptaddress]{revtex4-1}
\usepackage{amssymb,amsmath,amsfonts,bm}
\usepackage{graphicx,subfigure}
\usepackage{hyperref}
\usepackage{url}
\usepackage[dvipsnames]{xcolor}
\usepackage{float}
\begin{document}
\title{Study of Domains in the Ground State of the Two Dimensional Coulomb Glass}
\author{Preeti Bhandari}
\affiliation{Department of Physics, Jamia Millia Islamia, New Delhi 110025, India}
\author{Vikas Malik}\email{vikasm76@gmail.com}
\affiliation{Department of Physics and Material Science, Jaypee Institute of Information Technology, Uttar Pradesh, India}
\date{\today}

\begin{abstract}
We have annealed two dimensional lattice model of Coulomb glass using Monte Carlo simulations to obtain the minimum energy state (referred to as ground state). We have shown that the energy required to create a domain of linear size L in d dimensions is proportional to $L^{d-1}$. Using Imry-Ma arguments given for RFIM, one gets critical dimension $d_{c}\geq 2$ for Coulomb glass. The investigations in the transition region shows that the domain wall of the metastable state in the charge-ordered phase shifts as disorder is increased to give disordered ground state at higher disorder strength indicating phase coexistence. This coupled with discontinuity in magnetization is an indication of first-order type transition from charge-ordered phase to disordered phase. The structure and nature of Random field fluctuations of the domain in Coulomb glass are inconsistent with the assumptions of Imry and Ma as was also reported for RFIM. 

\end{abstract}
\maketitle

The Coulomb glass (CG) is a system in which all electron states are localised and they interact via long-range Coulomb potential. At low temperature, these localised electrons are unable to screen the Coulomb interactions effectively. The long range nature of the Coulomb interactions leads to a soft gap \cite{Pollak,Sri,Shk} in the single-particle density of states. This effect changes the conductivity from $ln\sigma \sim T^{1/4}$ to $T^{1/2}$ as temperature ($T$) is decreased \cite{Mott1,Mott2}. The formation of gap and the crossover of $T^{1/4}$ to $T^{1/2}$ at low T have been confirmed experimentally and numerically \cite{Mob}. Another important effect of Coulomb interaction is correlation effects, i.e. existence of collective hops instead of single electron hops \cite{Kno}. In recent years, focus has shifted from higher disorder to low disorder region \cite{Surer,Mobi}. It has been shown \cite{Goethe} numerically that in three-dimensional (3d) CG, the transition from fluid to the charge-ordered phase (COP) is consistent with the random field Ising Model (RFIM) university class. Whether the same is true in two-dimensional (2d) CG is yet to be investigated .\\
 \hspace*{3mm} The motivation of this paper is to understand the importance of Coulomb interactions in domain formation and how the structure of the domain differs from the short range model i.e. RFIM. The Imry-Ma arguments \cite{Imry} on which the initial theoretical papers on RFIM \cite{Vill,Grin,Stauffer,Fisher} were based, suggested that the energy required for the formation of a domain of linear size L in d-dimensions is $\mathcal{O}(L^{d-1})$. The amount of energy gained from the fluctuations of random field (RF) in the domain is $\mathcal{O}(L^{d/2})$, so the long range order will get destroyed for $d<2$. The ground state of 3d RFIM shows a transition from ferromagnetic to disordered state as disorder is increased \cite{Ogiel}. Binder \cite{Binder} argued that roughening of domain walls would stabilize the domain in two-dimensions and lead to destruction of ferromagnetic ordering. A rigorous proof was then given by Aizenman and Wehr \cite{Aizenman} stating that there is no long-range order in 2d RFIM. These arguments led to a critical dimension $d_{c}=2$.\\
\hspace*{3mm} Numerical evidence \cite{Seppala,Seppala2} shows roughening of domain walls and the ground state breaking into domains above a length scale that depends exponentially on the random field strength squared, further strengthened the argument that $d_{c}=2$. Experiments on 2d dilute antiferromagnets, showed that no long-range ordering is present \cite{RJ2,IB1}, but a possibility of first order transition in 3d has been observed \cite{Cow}.\\ 
\hspace*{3mm} Contradicting all the above work, evidence of numerical signs of transition in 2d RFIM at $T=0$ below a critical disorder was shown by Frontera and Vives \cite{Frontera}. In a seminar \cite{Aizen} in 2012, Aizenman also claims that the 2d RFIM exhibits a phase transition. In 2013 Sinha and Mandal \cite{Suman} used Monte Carlo simulations to show that for weak fields 2D RFIM possesses long-range ordering. The validity of the Imry-Ma arguments was tested by doing numerical calculations. The properties of domains were significantly different from the assumptions made by Imry and Ma \cite{Cambier,Esser}.\\
\hspace*{3mm} In this letter, we investigate the possibility of transition from charge-ordered phase (COP) to disordered phase (DP) and the properties of domain structure in the ground state via Monte Carlo (MC) annealing of the two-dimensional (2d) CG lattice model with on-site disorder \cite{Ef,Davies}. Our results are as follows:(i) We found indication of a first-order type transtion from COP to DP as seen in 2d RFIM \cite{Seppala2}. The ground state in DP consists of two large inter-penetrating domains. (ii) Although the long-range interactions in the system remain unscreened, the interaction energy of the domain is still $\mathcal{O}(L^{d-1})$ which allows one to use the Imry-Ma argument.  
(iii) The domain wall of the metastable state in the COP at $W_{c}^{-}$ shifts  to give disordered ground state at $W_{c}^{+}$ ($W_{c}$ is the critical disorder). (iv) In the disordered phase ($W_{c}^{+}$), our results shows that the domain structure and the nature of random-field (RF) fluctuations in the domain contradict the Imry-Ma assumptions but are consistent with the numerical work on RFIM \cite{Cambier}. Further extension of our work will be to see how the two large clusters formed at $W_{c}^{+}$ evolve as one increases the disorder. This will have interesting effects on the conductivity as well as on the nature of the phase \cite{DR,Vak}. Our method can be used to study long-range Ising systems at $T=0$. \\
\hspace*{3mm} We consider the classical 2d CG lattice model, in which the electron states are assumed to be localized around the sites of a regular lattice with lattice spacing $a\equiv 1$. We work with a case of half filling which implies that the number of electrons are half the total number of sites (N). We use the pseudospin variables $S_{i}=n_{i}-1/2$ where $n_{i}\in {0,1}$ is the occupation number at site i. The Hamiltonian of the system can now be written in spin language as 
\begin{equation}
H = \sum_{i} \phi_{i}S_{i} + \frac{1}{2} \sum_{i\neq j} J_{ij} S_{i}S_{j}
\end{equation}
where the unscreened Coulomb interactions are described as $J_{ij}=e^{2}/\kappa R_{ij}$, $\kappa$ is the dielectric constant and $R_{ij}$ is the distance between site i and j. We are using periodic boundary condition (PBC) and minimum-image convention for calculating $R_{ij}$ . $\phi_{i}$'s denote the random on-site energies, chosen randomly from a box distribution with interval [-W/2,W/2].  The particle-hole symmetry with symmetric disorder distribution lead to $\mu=0$. All the energies were measured in the unit of $e^{2}/\kappa a$.\\
\hspace*{3mm} We are here proposing an argument to calculate the energy of a regular domain (which is square for $d=2$ and cube for d=3) created in the ground state of a d-dimensional CG lattice model at half filling. The Hamiltonian of the system in terms of Hartree energy ($\varepsilon_{i}$) can be written as $H = \frac{1}{2} \sum^{N}_{i=1} (\varepsilon_{i} +2\phi_{i}) S_{i} $ where $\varepsilon_{i} = \sum_{i\neq j} J_{ij} S_{j}$. In the zero disorder limit, the ground state of a CG system has Antiferromagnetic ordering. So Hartree energy at each site is equal to d-dimension Madelung energy ($\varepsilon_{d}$). Staggered magnetisation defined as $\sigma=1/N \sum^{N}_{i=1}\sigma_{i}$, is the order parameter where $\sigma_{i}=(-1)^{i} S_{i}$. Cluster of nearest -neighbour sites which have same $\sigma_{i}$ is defined as a domain. As the system has anti-ferromagnetic ordering, each row on the lattice is charge neutral. For any charge, the contribution to its Hartree energy can be divided into two parts (a) from charges on the line (plane) on which the site is located (b) charges on few rows (planes) just above and below the charge under consideration and negligible contribution coming from rest of the lines(planes) \cite{Tasker}. There is no surface effect as we are using PBC. Now if we consider a large regular domain then the Hartree energy of the sites inside the domain will be equal to $\varepsilon_{d}$ using the reasoning given above. Extending the same argument, the Hartree energy of the site on the domain wall becomes approximately equal to $d-1$ Madelung energy ($\varepsilon_{d-1}$). This is because, for a site $i$ on the domain wall, $\varepsilon_{i}=\sum J^{F}_{ij} \sigma^{out}_{j} + \sum J^{F}_{ij} \sigma^{in}_{j} + \sum J^{F}_{ij} \sigma^{wall}_{j}$ where $J^{F}_{ij} = J_{ij} (-1)^{i+j}$ and $\sigma^{out}_{j},\sigma^{in}_{j},\sigma^{wall}_{j}$ describes $\sigma$ of sites outside the domain wall, inside the domain wall and on the domain wall respectively. The first two terms in the summation cancels out because $\sigma^{in}_{j}=- \sigma^{out}_{j}$. The third term is $\sim \varepsilon_{d-1}$ for a large domain. So the energy required to create a domain wall is $((\varepsilon_{d}-\varepsilon_{d-1})/2)\times P$ where P is proportional to $L^{d-1}$. Since energy gained from RF is still $\mathcal{O}(L^{d/2})$, Imry-Ma argument can be applied to a regular domain. This explains why 3d CG is in the same university class of RFIM as claimed earlier \cite{Goethe}. Whether long range order can exist for 2d CG is now a matter of further investigations.\\
\begin{figure}
\centering
\includegraphics[width=7.5cm,height=4cm]{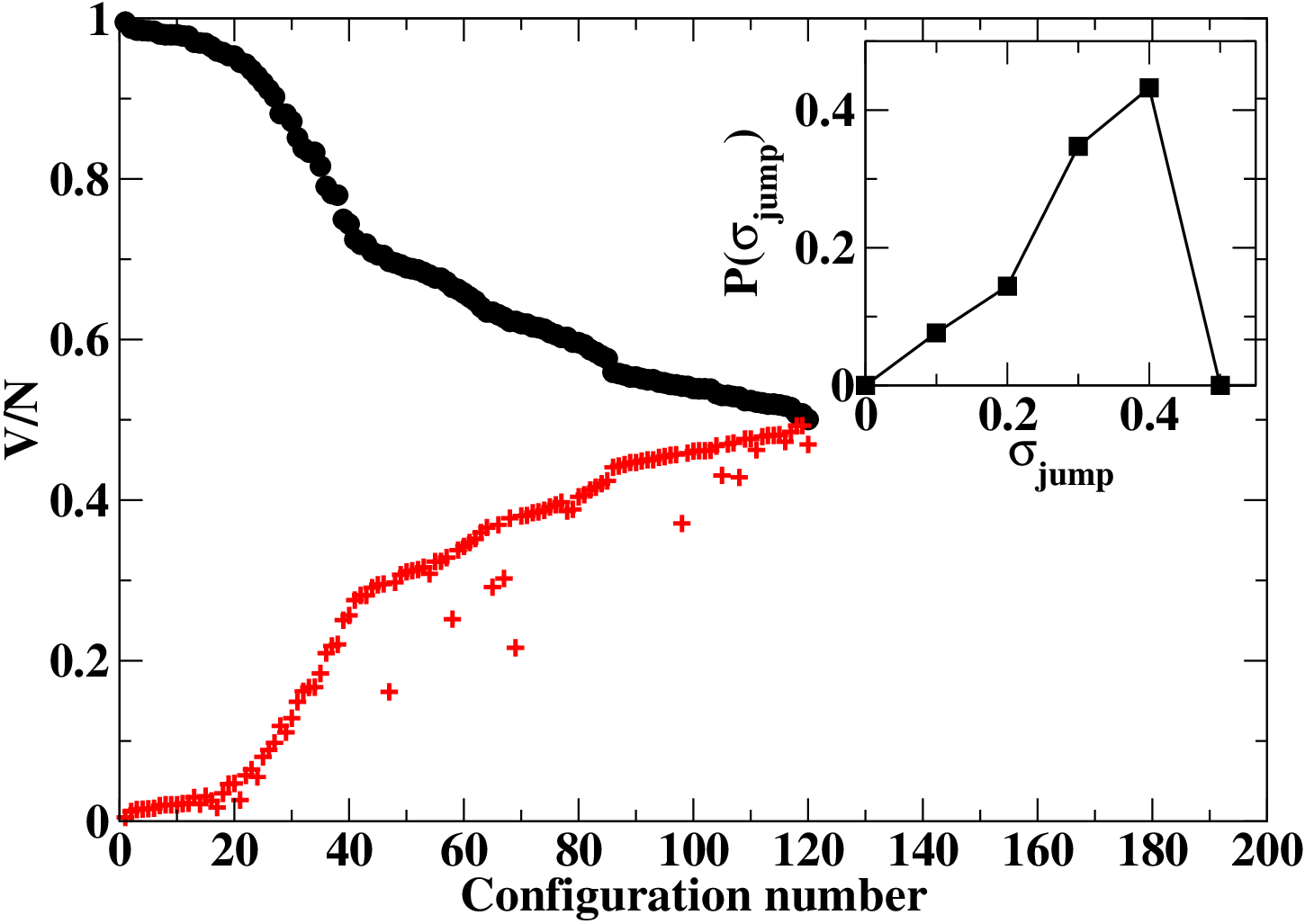}
\caption{\label{Fig1} (Color online) Largest ($\circ$) and second largest ($+$) domain size ($V$) in the DP ($W_{c}^{+}$) divided by the system size $N=64^{2}$. The largest domains are sorted in descending order. Inset shows the largest jump in $\sigma$ at each configuration.}
\end{figure}
\hspace*{3mm} We have done simulated annealing using Monte Carlo (MC) technique for $32\times 32$ and $64\times 64$ system. Kawasaki Dynamics was used as the number of electrons are conserved. A single Monte Carlo Step (MCS) involves randomly choosing two site of opposite spins for spin exchange (single electron hop). The initial system was completely random spin configuration $\lbrace S_{i}\rbrace$ with  half sites assigned with $S_{i}=1/2$ and the other half with $S_{i}=-1/2$. Annealing using Metropolis algorithm  \cite{Metro} was done from $T=1$ to $T=0.01$. Our longest run was $5\times10^{5}$ MCS per site at $T=0.01$. The investigations were carried out for different disorders strengths ($W=0.0$ to $0.40$). 
For each configuration, the sign of $\lbrace \phi_{i}\rbrace$ at each site were fixed and the strength of the randomness ($W$) was increased \cite{Seppala2,Esser}. This approach has an advantage that one is able to see the evolution of the metastable state of COP  to the ground state of DP as W is increased. After annealing was completed, we found system consisted of mostly two large domains and few small domains. The domains were identified using Hoshen-Kopelman algorithm \cite{Hoshen}. We then calculated the domain-domain interactions and found it to be negligible. This allowed us to flip the domains one by one to get the minimum energy state (ground state). The ground state was found to be stable against single electron hop. We then carried out the simulation using different $\lbrace S_{i}\rbrace$ but same $\lbrace \phi_{i}\rbrace$. We got domains with same structure which were pinned at a certain location for all $\lbrace S_{i}\rbrace$. Thus these domains belong to the same valley. Although our method does not find the exact ground state, it correctly identifies two valleys, one centred around $\mid \sigma \mid=0.5$ (COP) and other around $\mid \sigma \mid \neq 0.0$ (small). \\
\begin{figure}
\centering
\includegraphics[width=7.5cm,height=5.5cm]{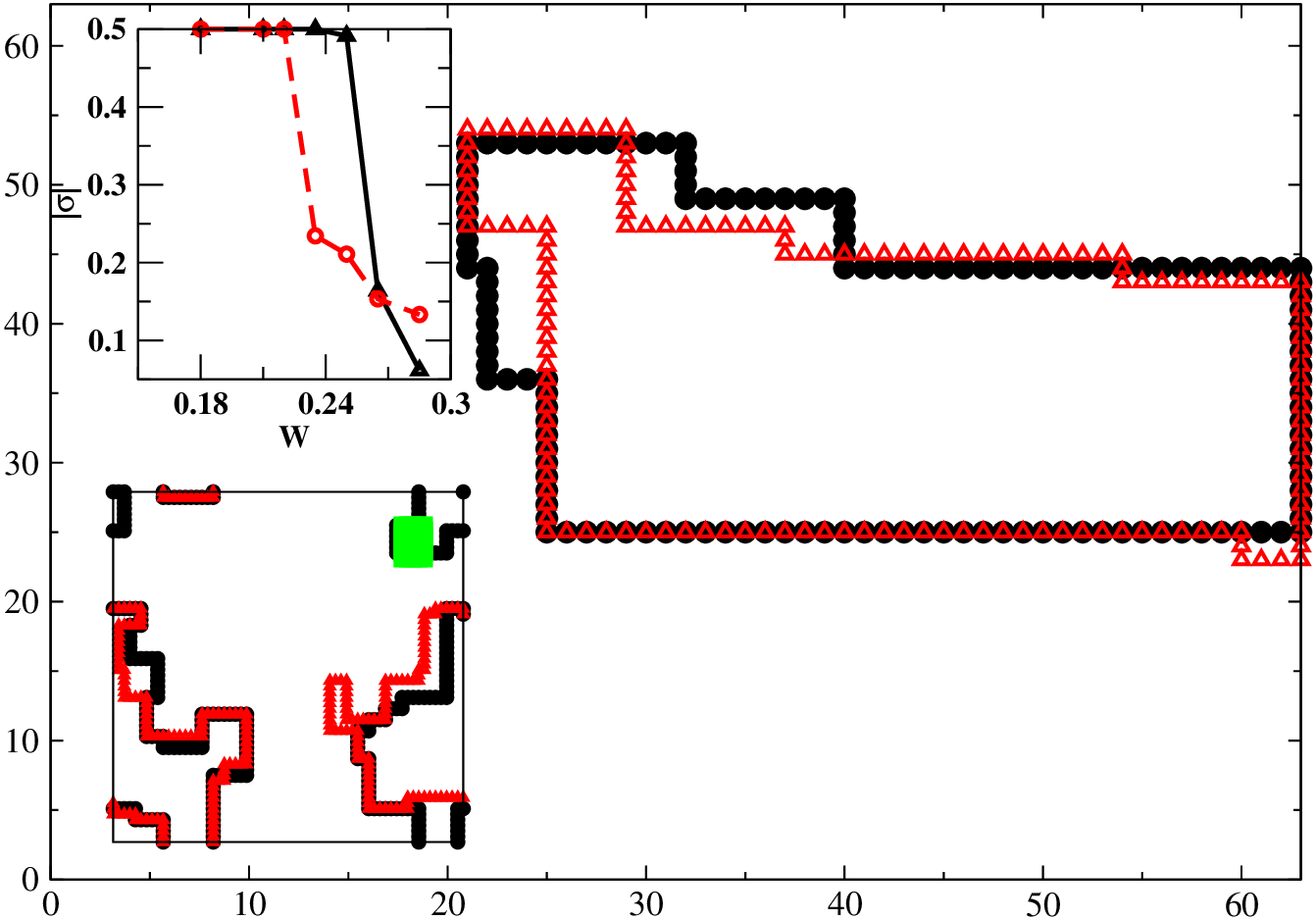}
\caption{\label{Fig2} (Color online) (a) Domain wall of Metastable state ($\bullet$) at $W=0.22$ and domain wall of ground state ($\bigtriangleup$) at $W=0.235$ are plotted for one of the configurations for $L=64$. (b) In the inset (bottom) domain Wall of metastable state ($\bullet$) at $W=0.235$ , ground state ($\blacksquare$) at $W=0.25$ and ground state ($\bigtriangleup$) at $W=0.265$. In the inset (top) Disorder dependence of $\mid \sigma \mid$ for (a) (dashed line) and (b) (solid line).}
\end{figure}
\hspace*{3mm} The investigations in this paper were done using the ground state at $W_{c}^{+}$ and the metastable state of the COP at $W_{c}^{-}$ (for each disorder configuration $W_{c}$ will be different). FIG. \ref{Fig1} shows the size of the largest and the second largest domains (which is non-zero for all configurations) in the ground state. For most of the configurations, COP breaks into two large domains as we move from $W_{c}^{-}$ (where $\sigma=\pm 0.5$) to $W_{c}^{+}$ (where $\sigma = small$) which results into discontinuity in staggered magnetisation at each configuration of disorder. So in the transition region the two phases coexist, which is a characteristic feature of a first-order transition. In simulations of 3d RFIM at finite T \cite{Reiger} and $T=0$ \cite{Ogiel,middle} small value of magnetization exponent indicative of discontinuity in magnetization at the transition was found but other parameters rule out the possibility of a first order transition. To further prove coexistence, we then compared the domains formed in the ground state at $W_{c}^{+}$ with the domains of the metastable state at $W_{c}^{-}$. From FIG \ref{Fig2}, one can see that the domain wall of the metastable state shifts to give the ground state as W was increased slightly. This shows that free energy which is equal to energy at $T=0$ has two minimas (valley) centred at $\mid \sigma \mid=0.5$ and $\mid \sigma \mid \approx small$, indicative of first order transition. From FIG. \ref{Fig1}, one can see that in few configurations, size of the second largest cluster is very small. The domain formation in one of such case is shown in the inset of FIG. \ref{Fig2}. One sees that a small jump is followed by a large jump in $\sigma$ which is the typical behaviour in such cases. This is confirmed by the probability distribution for the largest jump in $\sigma$ at each configuration shown in the inset of FIG. \ref{Fig1}. Such small jumps in magnetization and energy was also observed in 3d RFIM \cite{machta1,machta2} which not a singular behaviour. Hence to look at thermodynamically favourable transitions one should analyze large jumps. Thus the picture of a transition from valley centred around $\mid \sigma \mid=0.5$ to $\mid \sigma \mid \approx small$ valley is preserved. To support our argument that the transition is first order, we have done finite size scaling. Also the distribution of $\sigma$ around transition showing three peak structure is another evidence of coexisting phases (figures not shown here, for details refer \cite{Preeti}).\\
\hspace*{3mm} To test the validity of the Imry-Ma arguments on CG model, we focussed on the structure and the nature of RF fluctuations of the domains in the ground state. The compactness of the domains was checked by using a the power law relation \cite{Cambier} $S\thickapprox V^{\tau}$, where the surface (S) of the domain denote the number of sites on the domain wall and the volume (V) of the domain is the total number of sites in the domain. The value of the surface exponent $\tau$ for a compact domain is $1-1/d$.  FIG \ref{Fig4}  proves the validity of the relation for the CG system. The high value of $\tau$ indicates that the domains are non-compact.\\
\begin{figure}
\centering
\includegraphics[width=7.5cm,height=4cm]{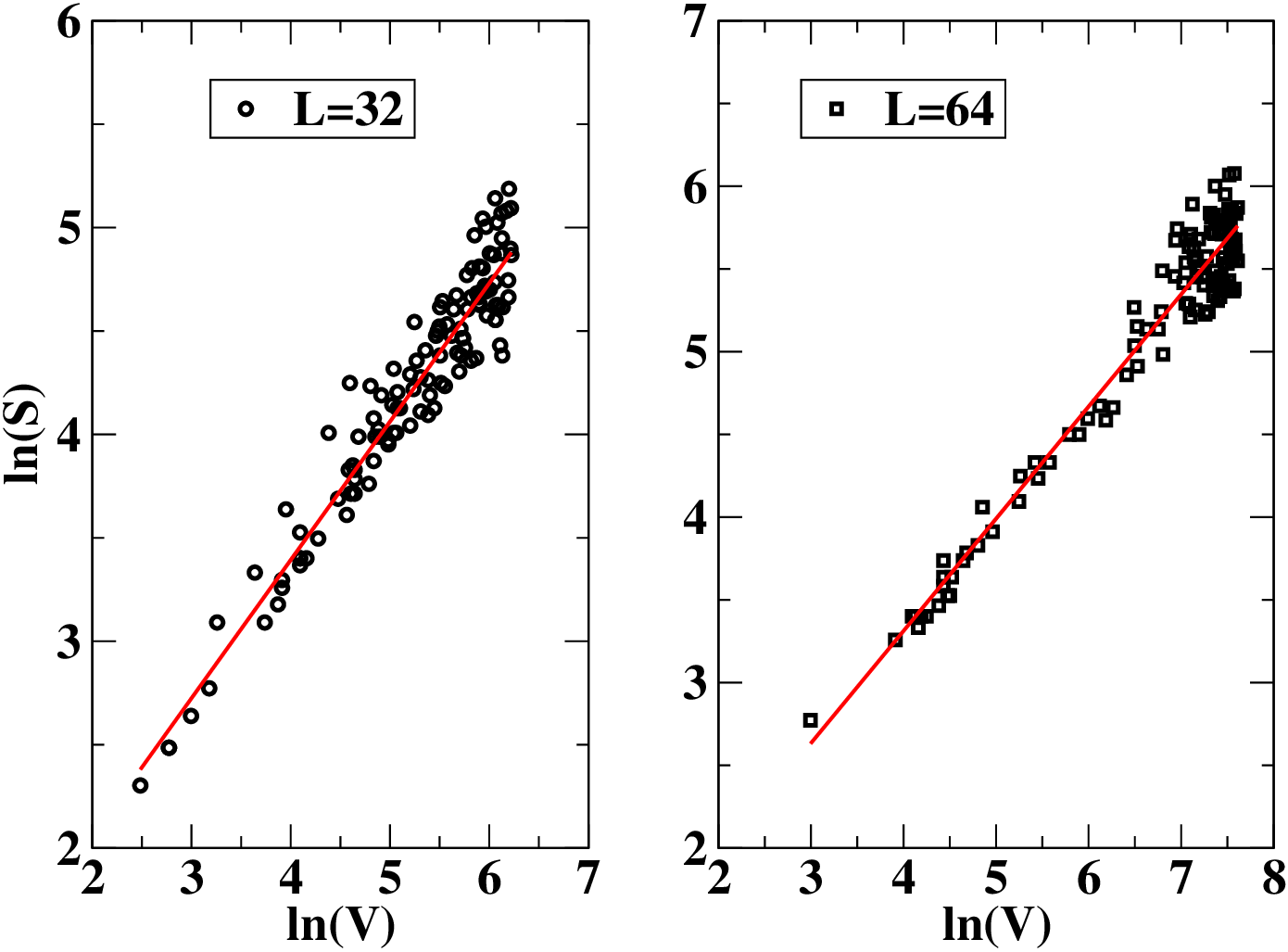}
\caption{\label{Fig4} (Color online) Logarithmic plot of surface vs size for the determination of surface exponent $\tau$ for 2d CG. $\tau = 0.6686$ and $0.6756$ for $L=32$ and $64$ respectively. }
\end{figure}   
\hspace*{3mm} Our theoretical argument for $DE \propto P$ (where DE is the domain energy) can be easily extended to a compact domains but needs to be numerically tested for non-compact domains. The plot of DE vs P (FIG. \ref{Fig5}) inset shows that the relation $DE = \eta \times P$ (where $\eta=0.033$) is valid for both system sizes. The value of $\eta$ calculated numerically is slightly higher then the predicted theoretical value for 2d which is $\eta= (\varepsilon_{2d}- \varepsilon_{1d})/2 \approx 0.0285$.  To understand this agreement in $\eta$ values, we have plotted the distribution of Hartree energies of the sites on the domain wall ($\varepsilon^{wall}_{i}$) and inside the domain wall ($\varepsilon^{inside}_{i}$) in FIG. \ref{Fig5}. The Hartree energies of the sites on the wall and inside the wall are distributed symmetrically around $\varepsilon_{1d}$ and $\varepsilon_{2d}$ respectively. This is the reason why our numerical results matches our argument given for DE calculation. Next we tested the hypothesis that the total random-field fluctuations (F) in a domain is typically a rms deviation and is proportional to the square root of $V$. A general power law expression \cite{Cambier} can be written as $F\approx V^{\lambda}$ where $\lambda$ was considered as an undetermined exponent.

\begin{figure}
\centering
\includegraphics[width=8cm,height=5cm]{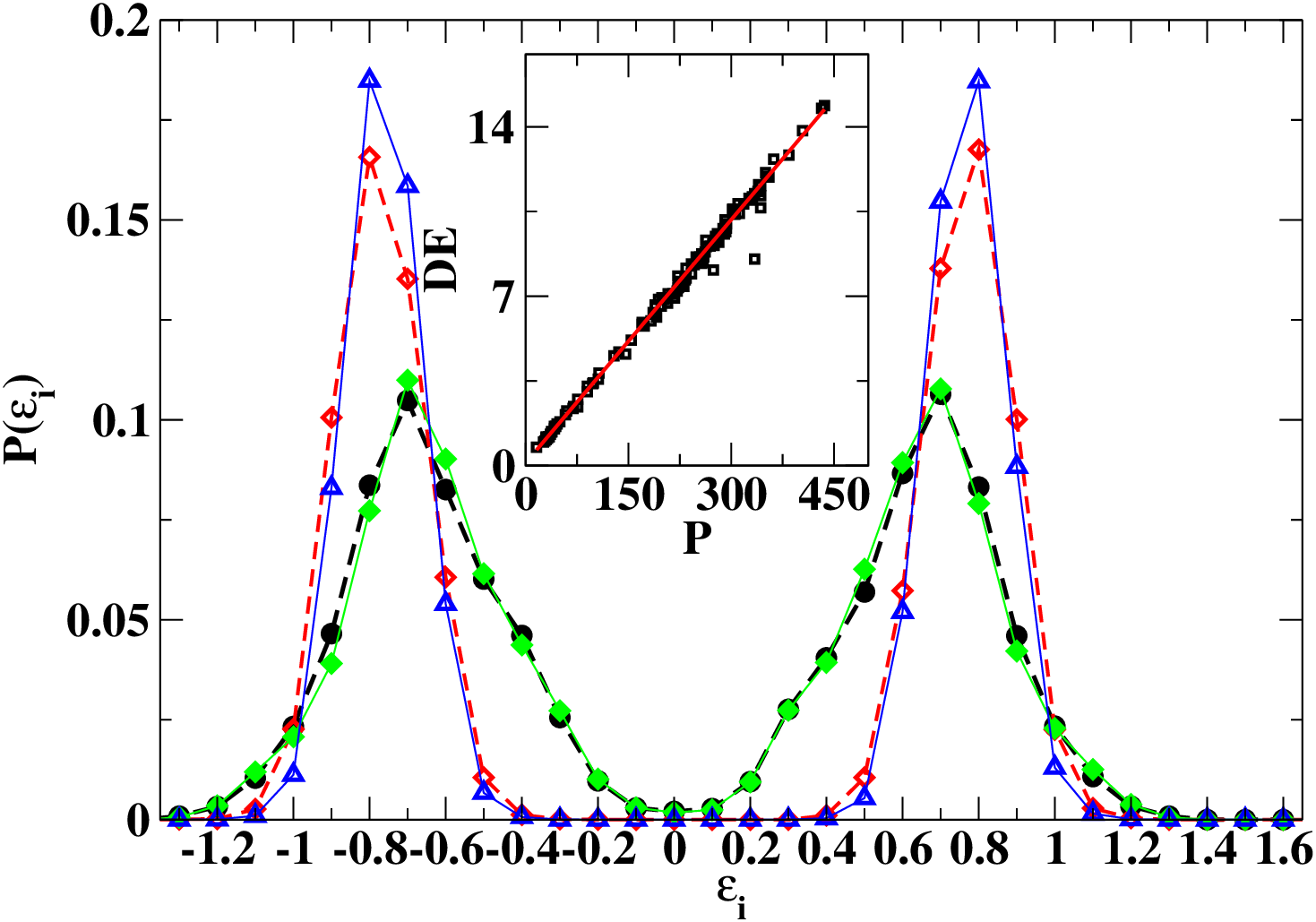}
\caption{\label{Fig5} (Color online) Distribution of Hartree energy of the sites on the domain wall ($\varepsilon_{i}^{wall}$ for $L=32$ ($\bullet$) and $L=64$ ({\color{Green}$\blacksquare$})) and on sites inside the domain wall where domain wall sites are excluded ($\varepsilon_{i}^{inside}$ for $L=32$ ({\color{Red}$\diamond$}) and $L=64$ ({\color{Blue}$\bigtriangleup$})). Inset shows the domain energy vs perimeter of the domain. $\eta = 0.0335$ for $L=64$.}
\end{figure}

\begin{figure}
\centering
\vspace{1mm}
\subfigure[]{\includegraphics[width=7cm,height=5cm]{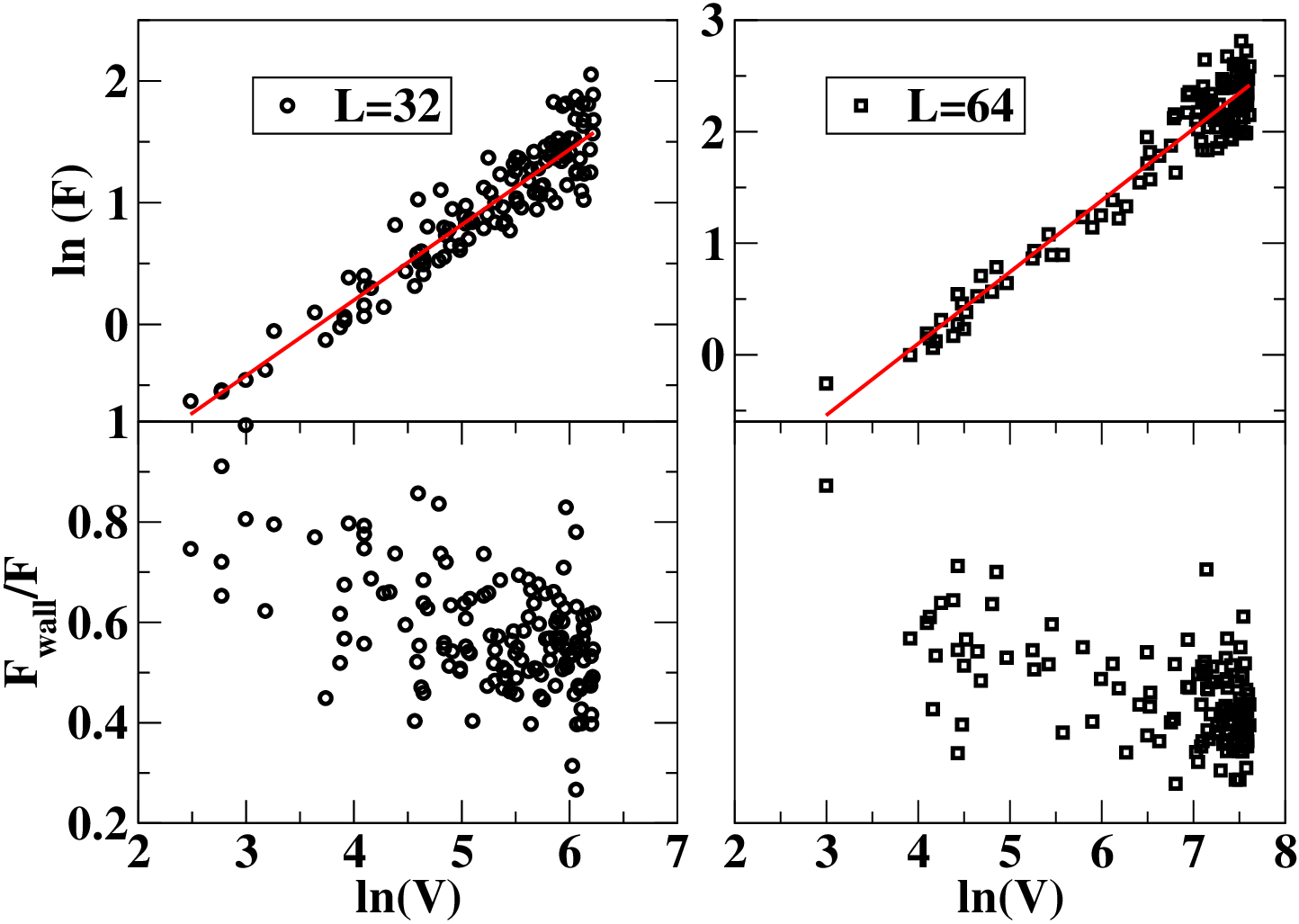}}
\subfigure{\includegraphics[width=7cm,height=3cm]{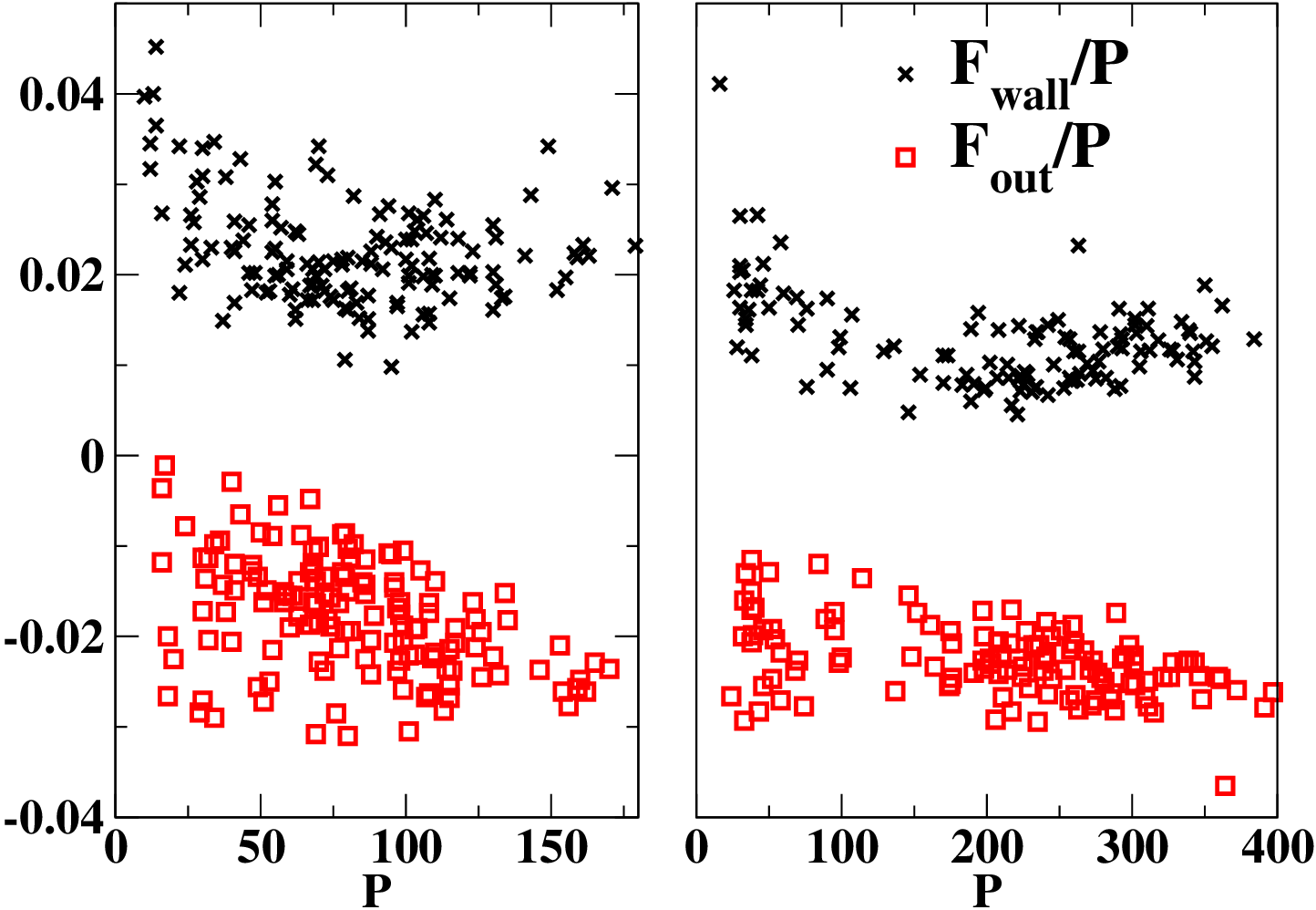}}
\caption{\label{Fig8} (Color online). Top: Logarithmic plot of RF fluctuation(F) vs volume(V) of the domains in ground state. The power law relation $  F\approx V^{\lambda}$ holds, giving $\lambda=0.6192$ and $0.6415$ for $L=32$ and $64$ respectively. Center: Ratio $F_{wall}/F$, where $F$ is the total RF fluctuation in the domain at ground state.  Bottom:RF fluctuation at domain wall($F_{wall}$) and layer just outside the domain wall ($F_{out}$). The y coordinate is the ratio $F/P$, with P the perimeter length, and the x ordinate is the perimeter length P.  }
\end{figure}

This relation is verified from FIG. \ref{Fig8}(Top). This value of $\lambda$ is significantly higher than the theoretical value ($\lambda=1/2$) assumed in RFIM. We have also calculated the ratio $F_{wall}/F$. In FIG \ref{Fig8}(center) one can see that the range of the ratio is $40\%$ to $60\%$ for most of the configurations, indicating that the random field energy of the domain is contained more in the domain boundary. We then calculated the random-field fluctuation of the sites on the domain wall ($F_{wall}$) and of the sites which are just outside the domain wall ($F_{out}$). FIG. \ref{Fig8}(bottom) shows that the random-field fluctuations are proportional to the perimeter of the domain and not its square root as assumed in previous theories. We have also checked that the location of the domains is independent of the initial spin configuration chosen and is determined by the random-field  configurations. The strong RF fluctuations on the wall lead to the pinning of the domain wall which is the reason for the metastability of the domains. So our results suggests that the domains in the ground state are pinned, non-compact and the RF fluctuations are contained more at the domain boundary. These results are consistent with the numerical work \cite{Cambier,Esser} done on RFIM.\\
\hspace*{3mm} \textit{Conclusions -} We have shown that the Imry-Ma argument for short range RFIM can be extended to CG system at half filling leading to $d_{c}=2$. To verify the argument, we numerically investigated the 2d CG lattice model using MC annealing. Our numerical work shows a phase transition from COP to disordered phase of first-order type. The transition is driven by the rearrangement of domain wall of the metastable state in COP as W is increased to give disordered phase. The domains formed are non-compact and the RF fluctuations are contained more at the domain wall, in contradiction with Imry-Ma assumptions.       \\
\hspace*{3mm} We thank late Professor Deepak Kumar for useful discussions on the subject. We wish to thank NMEICT cloud service provided by BAADAL team, cloud computing platform, IIT Delhi for the computational facility. Preeti Bhandari acknowledges UGC, Govt. of India for financial support through UGC-BSR fellowship (F.25-1/2013-14(BSR)/7-93/2007(BSR)).


\end{document}